\documentclass[conference]{IEEEtran}
\IEEEoverridecommandlockouts
\usepackage{cite}
\usepackage{amsmath,amssymb,amsfonts}
\usepackage{algorithmic}
\usepackage{graphicx}
\usepackage{textcomp}
\usepackage{xcolor}
\usepackage{balance}
\def\BibTeX{{\rm B\kern-.05em{\sc i\kern-.025em b}\kern-.08em
    T\kern-.1667em\lower.7ex\hbox{E}\kern-.125emX}}

\usepackage{comment}
\usepackage{xspace}
\usepackage{enumitem}
\usepackage[ruled,linesnumbered,noend]{algorithm2e}
\usepackage{subfigure}
\usepackage{xcolor}
\usepackage{multirow}
\usepackage{booktabs}
\usepackage{url}
\usepackage{amsthm}
\usepackage{makecell}
\usepackage{multicol}
\usepackage{xspace}

\newcommand{\method}{ASOC\xspace}
\newcommand{\adv}{ASOC+\xspace}
\newcommand{\deeprane}{DeepRaNE\xspace}
\newtheorem{problem}{Problem}

\newcommand{\red}[1]{\textcolor{red}{#1}}

\newcommand{\supplelink}{\url{https://github.com/jihoonko/ASOC}\xspace}
\newcolumntype{C}[1]{>{\centering\let\newline\\\arraybackslash\hspace{0pt}}m{#1}}

\begin{document}

\title{Deep-Learning-Based Precipitation Nowcasting with Ground Weather Station Data and Radar Data
}

\author{\IEEEauthorblockN{Jihoon Ko\IEEEauthorrefmark{1}, Kyuhan Lee\IEEEauthorrefmark{1}, Hyunjin Hwang, and Kijung Shin}
\IEEEauthorblockA{\textit{Kim Jaechul Graduate School of AI} \\
\textit{Korea Advanced Institute of Science and Technology}\\
Seoul, South Korea \\
\{jihoonko, kyuhan.lee, hyunjinhwang, kijungs\}@kaist.ac.kr}
}

\maketitle
\begingroup\renewcommand\thefootnote{\IEEEauthorrefmark{1}}
\footnotetext{Equal contribution.}
\endgroup

\begin{abstract}
Recently, many deep-learning techniques have been applied to various weather-related prediction tasks, including precipitation nowcasting (i.e., predicting precipitation levels and locations in the near future).
Most existing deep-learning-based approaches for precipitation nowcasting, however, consider only radar and/or satellite images as inputs, and meteorological observations collected from ground weather stations, which are sparsely located, are relatively unexplored.
In this paper, we propose \method, a novel attentive method for effectively exploiting %
ground-based meteorological observations from multiple weather stations.
\method is designed to capture  temporal dynamics of the observations and also contextual relationships between them.
\method is easily combined with existing image-based precipitation nowcasting models without changing their architectures. 
We show that such a combination improves the average critical success index (CSI) of predicting heavy (at least 10 mm/hr) and light (at least 1 mm/hr) rainfall events at 1-6 hr lead times by 5.7\%, compared to the original image-based model,
using the radar images and ground-based  observations around South Korea collected from 2014 to 2020.

\end{abstract}


\section{Introduction}
\label{sec:intro}

Recently, deep learning techniques, especially computer vision techniques, have been applied to forecasting various weather-related events, and such approaches \cite{shi2015convolutional,shi2017deep,ko2022effective,agrawal2019machine,sonderby2020metnet,lebedev2019precipitation} often outperform traditional methods in the field. 
A representative example is precipitation nowcasting, which is short-term (e.g., at $0$-$6$ hour lead times \cite{schmid2019nowcasting,wilson2004sydney}) location-specific forecasting of precipitation.
According to \cite{espeholt2021skillful}, current approaches equipped with deep convolutional networks outperform HRRR \cite{benjamin2016north}, which is one of the state-of-the-art numerical weather prediction  models, with lead times up to $12$ hours.

For precipitation nowcasting, U-Net \cite{ronneberger2015u} and ConvLSTM \cite{shi2015convolutional}, which were originally designed for semantic segmentation and spatio-temporal sequence forecast, respectively, have been used mainly as backbone network architectures. For example, Agrawal et al. \cite{agrawal2019machine} and Lebedev et al. \cite{lebedev2019precipitation} adapted U-Net and used radar-reflectivity images and satellite images as inputs.  Ko et al. \cite{ko2022effective} also adapted U-Net to demonstrate the effectiveness of their proposed training strategies for deep-learning-based precipitation nowcasting.
Shi et al. \cite{shi2015convolutional} demonstrated that ConvLSTM  outperforms optical-flow-based methods and fully-connected LSTM  on precipitation nowcasting. 
In order to improve the performance of precipitation nowcasting, ConvLSTM was extended to learn additional location-variant structures  \cite{shi2017deep}, and it was also extended with exponentially dilated convolution blocks, which enhance expressive power by capturing additional spatial information \cite{espeholt2021skillful}.

Most deep-learning-based approaches (e.g., \cite{agrawal2019machine,lebedev2019precipitation,ko2022effective,shi2015convolutional,shi2017deep}) consider only radar images and satellite images as inputs, and meteorological observations from ground weather stations have been underutilized. 
Although radar and satellite images, which are in a grid format, are naturally fed into deep convolutional neural networks (e.g., U-Net and ConvLSTM),
ground-based meteorological observations are not naturally represented in a grid format since ground weather stations are sparsely located.
Although interpolation techniques, such as Inverse Distance Weighting \cite{shepard1968two} and Kriging \cite{metheron1971theory}, can be used to resolve this issue, they are expensive both in time and memory, especially to obtain high-resolution data.
Thus, in order to utilize ground-based meteorological observations together with radar and satellite images, deep-learning models should be able to utilize input data of different formats efficiently and effectively.

In this paper, to address the aforementioned challenge, we propose Attentive Sparse Observation Combiner (\method), a novel deep-learning model for precipitation nowcasting based on meteorological observations collected from multiple ground weather stations. In a nutshell, \method combines LSTM \cite{hochreiter1997long} and Transformer \cite{vaswani2017attention} %
to capture %
temporal dynamics of the ground-based observations and also contextual relationships between them. Specifically, \method uses LSTM, which processes observations in chronological order, to capture temporal dynamics, and it uses Transformer-style attention blocks  between LSTM cells to capture contextual relationships between observations.
Another advantage of \method is that it is easily combined with existing image-based models, without any change in their design.
In our experiments, we use \adv, where \method is combined with \deeprane
\cite{ko2022effective}, which is one of the state-of-the-art image-based models.

We evaluate our approaches using radar-reflectivity images and ground-based observations (from $714$ weather stations) around South Korea collected for seven years (spec., from $2014$ to $2020$). We demonstrate that \adv improves the average critical success index (CSI) of predicting heavy ($\geq10$ mm/hr) and light ($\geq1$ mm/hr)  rainfall events at 1-6 hr lead times by $5.7$\%, compared to \deeprane.
For \textbf{reproducibility}, we made the source code used in the paper publicly available at \supplelink.

In Section~\ref{sec:related}, we briefly review related studies. %
In Section~\ref{sec:notation}, we introduce the notations used in this paper and define the precipitation nowcasting problem. In Section~\ref{sec:method}, we present \method and \adv. In Section~\ref{sec:exp}, we provide experimental results. Lastly, in Section \ref{sec:conclusion}, we provide conclusions.

\section{Related Work}
\label{sec:related}

In the machine-learning literature, precipitation nowcasting is often formulated as pixel-wise classification of precipitation levels in the near future from input radar-reflectivity images, and satellite images are often used additionally as inputs. %

Among convolutional neural networks (CNNs), U-Net \cite{ronneberger2015u} has been widely used for precipitation nowcasting \cite{agrawal2019machine, lebedev2019precipitation, ravuri2021skilful, ko2022effective}.
U-Net was originally designed for an image segmentation task, i.e., pixel-wise classification.
For example, Lebedev et al. \cite{lebedev2019precipitation} used U-Net for precipitation detection, which is formulated as a pixel-wise binary-classification problem.
Agrawal et al. \cite{agrawal2019machine} divided precipitation levels into four classes and used U-Net for pixel-wise multiclass classification. %
Based on a similar multiclass classification formulation, Ko et al. \cite{ko2022effective} proposed training strategies for precipitation nowcasting (spec., a pre-training scheme and a loss function) and demonstrated their effectiveness using a U-Net-based model.

Moreover, in order to aggregate both spatial and temporal information, there have been several attempts to combine recurrent neural networks (RNNs) (e.g., LSTM \cite{hochreiter1997long}) into CNNs \cite{shi2015convolutional, shi2017deep, sonderby2020metnet}. 
For example, Shi et al. \cite{shi2015convolutional} proposed ConvLSTM, which has
convolutional structures in the input-to-state and state-to-state transitions in LSTM.
Shi et al. \cite{shi2017deep} extended ConvLSTM to TrajGRU, which can learn location-variant connections between RNN layers.
Sønderby et al. \cite{sonderby2020metnet} proposed MetNet, which uses ConvLSTM as its temporal encoder and adapts axial attention structure for its spatial encoder.
Ravuri et al. \cite{ravuri2021skilful} pointed out that deep-learning based approaches tend to provide blurry predictions, especially at long lead times, %
and they used a conditional generative adversarial network \cite{mirza2014conditional} consists of ConvGRU-based \cite{ballas2016delving} generator and the spatial and temporal discriminators to address this limitation.
Espeholt et al. \cite{espeholt2021skillful} extended ConvLSTM with exponentially dilated convolution blocks, which enhance expressive power by capturing additional spatial information.

Several studies utilized meteorological observations from multiple weather stations as inputs to predict weather-related events.
For example, Seo et al. \cite{seo2018automatically} considered temperature forecasting. They generated a graph, where each node corresponds to a weather station, and inferred the data quality of each station, during training, by applying the graph convolutional network (GCN) to the generated graph.
Wang et al. \cite{wang2020nowcasting} focused on short-term intense precipitation (SIP) nowcasting.
They generated a graph and its feature, by identifying and clustering convective cells from
radar-reflectivity images, and used them, together with ground-based observations, as the inputs of a random forest classifier.
In contrast to our deep-learning-based approach, they did not employ any deep-learning techniques to process radar images and ground-based observations together.

\section{Basic Notations \& Problem Definition}
\label{sec:notation}
In this section, we introduce basic notations and formulate the precipitation nowcasting problem. 

\subsection{Basic Notations}
The frequently-used symbols are listed in Table~\ref{tab:notations}.
We use $R^{(t)}_x\in \mathbb{R}$ to indicate the radar reflectivity in dBZ at time $t$ in each region $x$, and we use $R^{(t)}$ to indicate the whole radar-reflectivity image at time $t$.
We use $I$ to denote the set of regions where ground weather stations are located.
Then, $O^{(t)}_x\in \mathbb{R}^{d}$ denotes the ground-based observations in each region $x\in I$ at time $t$, and $O^{(t)}$ denotes the ground-based observations at time $t$ from all regions in $I$.
Lastly, $C^{(t)}_x$ indicates the ground-truth precipitation class (see the following subsection for precipitation classes) in each region $x\in I$ at time $t$, and
$\hat{C}^{(t)}_x$ indicates the predicted probability distribution over all precipitation classes for each region $x$ at time $t$.

\begin{table}[t]
    \centering
    \caption{Frequently used notation.}
    \scalebox{0.95}{
    \begin{tabular}{r|l}
        \toprule
        Notation & Description \\
        \midrule
        $t$ & time (unit: minutes) \\
        \midrule
        $R^{(t)}_x\in \mathbb{R}$ & radar reflectivity at time $t$ in each region $x$ (unit: dbZ) \\
        \midrule
        $R^{(t)}$ & radar reflectivity image at time $t$ in all regions \\
        \midrule
        $I$ & set of regions where ground weather stations are located \\
        \midrule
        $O^{(t)}_x \in \mathbb{R}^{d}$ & ground-based observations at time $t$ in each region $x$  \\
        \midrule
        $O^{(t)}$ & ground-based observations at time $t$ in all regions \\
        \midrule
        $C^{(t)}_x$ & ground-truth precipitation class at time $t$ in each region $x$ \\
        \midrule
        $\hat{C}^{(t)}_x$ & predicted probability distribution over \\
        & precipitation classes at time $t$ in each region $x$ \\
        \bottomrule
    \end{tabular}
    }
    \label{tab:notations}
\end{table}

\subsection{Problem Definition}

The goal of  \emph{precipitation nowcasting}  is to predict precipitation levels and locations at very short lead times. In this paper, we formulate the problem as a location-wise classification problem, as in \cite{ko2022effective}. Specifically, we split precipitation levels into three classes: (a) \textsc{Heavy} for precipitation at least 10 mm/hr, (b) \textsc{Light} for precipitation at least 1 mm/hr but less than 10 mm/hr, and (c) \textsc{Others} for precipitation less than 1 mm/hr. 
We also frequently use a combined class named \textsc{Rain} (=\textsc{Heavy+Light}) for precipitation at least 1 mm/hr.
As inputs, we use ground-based observations collected from multiple weather stations and radar-reflectivity images collected for an hour.
We assume that both are collected every 10 minutes. For example, if we perform prediction at time $t$ in minutes, the inputs are (a) seven radar reflectivity images at times $\{t - 60, t-50, \cdots, t\}$, i.e., $R^{(t-60)}$, $R^{(t-50)}$, $\cdots$, $R^{(t)}$, and (b) seven snapshots of ground observations at times $\{t - 60, t-50, \cdots, t\}$, i.e., $O^{(t-60)}, O^{(t-50)}, \cdots, O^{(t)}$. Lastly, the range of lead times is chosen in 1-hour increments from a minimum of $1$ hour to a maximum of $6$ hours. In other words, a target time $t'$ in minutes is given among $\{t + 60, t + 120, \cdots, t + 360\}$ as an additional input.   
To sum up, the precipitation nowcasting problem considered in this work can be summarized as follows:

\begin{problem}\label{prob}
\textsc{\normalfont{(Precipitation Nowcasting at Time $t$)}} 
\begin{itemize}[leftmargin=*]
    \item \textbf{Given:} %
    \indent (1) a target time $t' \in \{t+60, t+120, \cdots, t+360\}$, \\
    (2) radar reflectivity images $R^{(t-60)}$, $R^{(t-50)}$, $\cdots$, $R^{(t)}$, and (3) ground-based observations $O^{(t-60)}, O^{(t-50)}, \cdots, O^{(t)}$
    \item \textbf{Find:} a prediction function $\Phi$
    \item \textbf{to Maximize:} classification performance.
\end{itemize} 
\end{problem}

In this work, we use the critical success index (CSI) and the F1 score to measure the performance. They are  suitable for class-imbalanced datasets and thus have been used widely for precipitation nowcasting \cite{agrawal2019machine,sonderby2020metnet,ko2022effective}. The scores are defined for each precipitation class $c$ (e.g., \textsc{Heavy} and \textsc{Rain}) as follows:

\begin{equation}
    \text{CSI}_{c} = \frac{\text{TP}_{c}}{\text{TP}_{c} + \text{FP}_{c} + \text{FN}_{c}}, \label{eq:csi}
\end{equation}
\begin{equation}
    \text{F1}_{c} = \frac{2 \cdot \text{TP}_{c}}{2 \cdot \text{TP}_{c} + \text{FP}_{c} + \text{FN}_{c}}, \label{eq:f1}
\end{equation}
where %
$\text{TP}_c$, $\text{FP}_c$, $\text{FN}_c$ stand for the counts of true positives, false positives, and false negatives, respectively, for each precipitation class $c$.

\section{Proposed Method: \method and \adv}
\label{sec:method}
\begin{figure*}
    \centering
    \includegraphics[width=\linewidth]{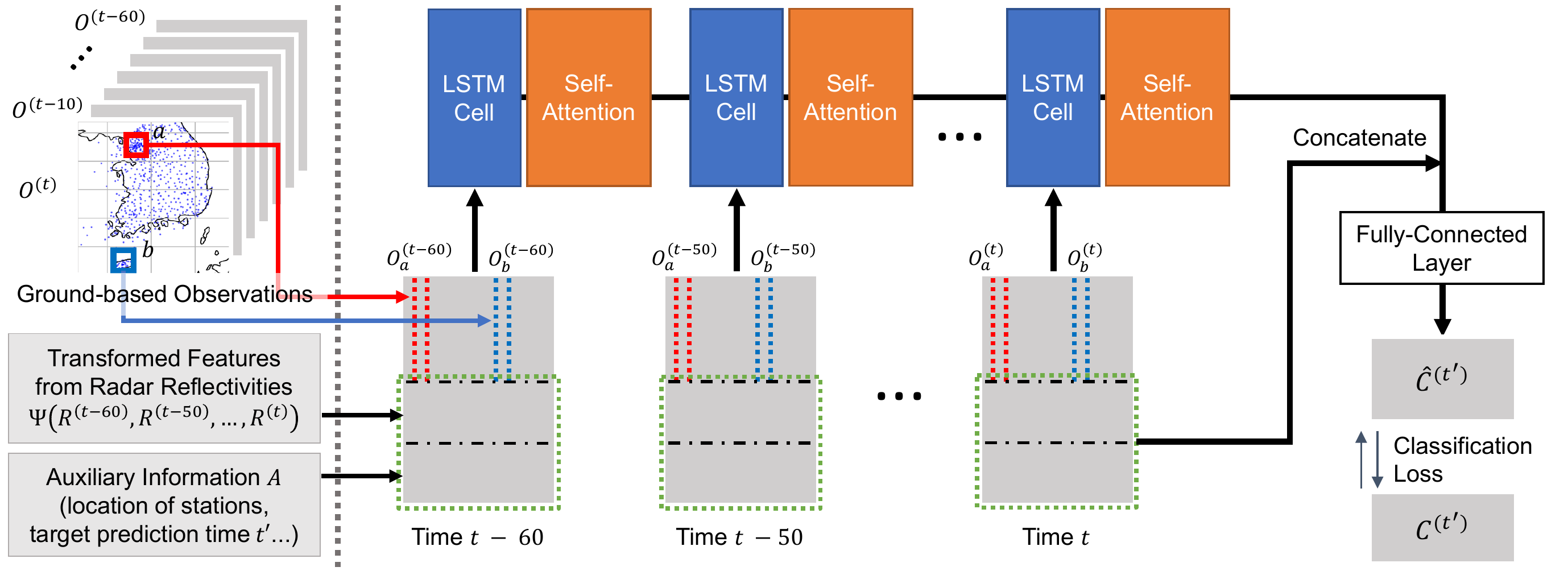}
    \caption{\underline{\smash{The overall inference process of \adv.}} Each ground-based observation snapshot is concatenated with (a) the pixel embeddings transformed from radar reflectivity images (e.g., by \deeprane \cite{ko2022effective}) and (b) auxiliary information. Then, the concatenated features are used as the input of a recurrent architecture that consists of LSTM cells and  self-attention modules. Lastly, the final prediction results are determined based on the output of the recurrent architecture.}
    \label{fig:overview}
\end{figure*}

In this section, we present \method, a novel deep-learning model for precipitation nowcasting based on meteorological  observations collected from multiple weather stations. We also discuss \adv, where \method is combined with a radar-image-based model for precipitation nowcasting.

\subsection{Overview}
In order to fully utilize multiple ground-based observations, we need to consider the following aspects:
\begin{itemize}[leftmargin=*]
    \item \textbf{Temporal dynamics of observations:} A sequence of ground-based observations over time are given as inputs, and temporal dynamics of them are useful for prediction.
    \item \textbf{Contextual relationships between observations:} 
    Ground-based observations collected from different weather stations are related to each other, and the strength of relation between each pair, which is useful for prediction, depends on context, such as lead times, overall weather conditions, and distance between weather stations. %

\end{itemize}

In order to capture the temporal dynamics of ground-based observations, we parameterize the function $\Phi$ in Problem \ref{prob} using a recurrent neural network (RNN), specifically LSTM \cite{hochreiter1997long}. Additionally, %
we use self-attention \cite{vaswani2017attention} to learn contextual relationships between pairs of ground weather stations. %
We provide a pictorial description of the overall architecture of \adv in Figure \ref{fig:overview}, and we describe the architecture and the inference process in detail in the following subsections. 

\subsection{Exploiting temporal dynamics}
Given consecutive snapshots of ground-based observations over time, capturing temporal dynamics of them is vital to precipitation nowcasting.
Instead of processing the observations using a permutation-invariant model, which results in loss of temporal information,
it is reasonable to use recurrent architectures for the parameterization and feed the inputs into the model in sequential order.
We use LSTM \cite{hochreiter1997long}, one of the most popular ones among such models. Note that, each LSTM cell processes inputs for one weather station at a time.
Thus, inside each LSTM cell, the observations obtained from one station do not directly affect the outputs for the other stations.

\subsection{Exploiting contextual relationships between observations}
Instead of considering  relation between observations inside the LSTM cells, we consider them separately in the self-attention blocks between LSTM cells. Specifically, among the hidden states and cell states of the LSTM, we only transform the hidden states using an attention mechanism while keeping the cell states as they are. For self-attention, we used an encoder layer of Transformer \cite{vaswani2017attention}, which consists of a multi-head attention layer and a feed-forward network. 
In our experiments,
we did not use layer normalization at all since the input ground-based  observations  contain scale-sensitive features, such as rainfall amount. 

\begin{algorithm}[t]
\caption{Inference step of \adv}\label{alg:overview}
\SetKwInput{KwInput}{Input}
\SetKwInput{KwOutput}{Output}
\SetKw{KwBy}{by}
\SetKwComment{Comment}{// }{}
\KwInput{(a) time of prediction $t$ \\
(b) target time $t' \in \{t + 60, \cdots, t + 360\}$ \\
(c) snapshots of ground observations $O^{(t-60)}$, $\cdots$, $O^{(t)}$ \\
(d) snapshots of radar reflectivities $R^{(t-60)}$, $\cdots$, $R^{(t)}$\\
(e) auxiliary information of all regions $\{A_i\}_{i \in I}$\\
(f) trained radar-image-based model $\Psi$ (spec., \cite{ko2022effective}) \\
(g) trained parameters of \adv: $\mathbf{W}$, $\mathbf{b}$, and the parameters of $\text{\normalfont{LSTMCell}}$ and  $\text{\normalfont{MultiHeadAttention}}$
}
\KwOutput{Predicted probability distributions $\{\hat{C}_i^{(t)}\}_{i\in I}$}

$\tilde{R}$ $\gets$ $\Psi(R^{(t-60)}, \cdots, R^{(t)}, t' - t)$ \\
$S_{h}$ $\gets$ $\{\emptyset\}_{i \in I}$ \Comment*[f]{hidden state of LSTM} \\
$S_{c}$ $\gets$ $\{\emptyset\}_{i \in I}$ \Comment*[f]{cell state of LSTM}\\
\For{$j \gets t-60$ \KwTo $t$ \KwBy $10$}{
    \ForEach{$i \in I$}{
        $X^{(j)}_i$ $\gets$ $\text{\normalfont{CONCATENATE}}(O^{(j)}_i, \tilde{R}_i, A_i)$ \\
        $(S_{h,i}, S_{c,i})$ $\gets$ $\text{\normalfont{LSTMCell}}(X^{(j)}_i, S_{h,i}, S_{c,i})$ \\
    }
    $\{S_{h, i}\}_{i \in I}$ $\gets$ $\text{\normalfont{MultiHeadAttention}}(\{S_{h, i}\}_{i \in I})$ \\
}
\ForEach{$i \in I$}{
    $Y_{i}$ $\gets$ $\text{\normalfont{CONCATENATE}}(S_{h,i}, \tilde{R}_i, A_i)$ \\
    $\hat{C}^{(t')}_{i}$ $\gets$ $\mathbf{W}{Y_i} + \mathbf{b}$
}
\textbf{return} $\{\hat{C}_i^{(t')}\}_{i\in I}$
\end{algorithm}

\subsection{Integration to  image-based prediction models}
\label{sec:method:integration}
In addition to ground-based observations,
any extra information (e.g., radar reflectivity in each region) can be fed into our model.
As auxiliary information, we use the location of each station and the lead time. Specifically, we use (a) a two-dimensional vector representing the location of each region, (b) a two-dimensional vector representing the observation date, (c) a two-dimensional vector representing the observation time, and (d) a six-dimensional one-hot vector for the lead time. Note that the last two vectors are the same for all regions.
For each two-dimensional vector for location, we use the coordinates after min-max normalization.\footnote{$x_{scaled} = \frac{2x -lb -ub}{ub - lb}$, where $x$ is an input feature value, and $ub$ and $lb$ are the $95$-th percentile and $5$-th percentile of x, respectively.} For the two-dimensional vectors for observation date and time, we perform min-max normalization and feed the result into the sine and cosine functions (see \cite{petnehazi2019recurrent} for details of the positional encoding method).
The vectors are all concatenated to the vector of ground-based observations at each weather station.

Moreover, output pixel embeddings of image-based precipitation nowcasting models can also be used as an additional input.
While any existing image-based precipitation nowcasting model can be used for this purpose,
we choose \deeprane \cite{ko2022effective}, which is a state-of-the-art radar-image-based prediction model to be combined with \method. We call the combined model \adv. To sum up, for each region $x \in I$,  the final input feature of \adv for region $x$ and time $t$ is the concatenated vector of (a) the ground observation feature $O_x^{(t)}$, (b) the pixel embedding of the corresponding region obtained by \deeprane, and (c) the two-dimensional vector for representing the location of $x$, and (d) the ten-dimensional common auxiliary information vector.
In \method, all vectors except for (b) are concatenated to be used as the input.

After obtaining the output of the LSTM part, \method and \adv obtain the final output probability distribution $\{\hat{C}^{(t)}_i\}_{i \in I}$ for each region through an additional fully-connected layer. 
Together with the output of the LSTM part,
the final input vector of the LSTM part are also fed into the last fully-connected layer. The pseudocode for the overall inference step of \adv is provided in Algorithm \ref{alg:overview}. 

\subsection{Training method}
For training \adv, we adapt the training protocol in \cite{ko2022effective}. Specifically, we split the training process into two steps: (a) pre-training with radar images and (b) fine-tuning with radar images and ground-based observations. During pre-training, we only consider the parameters of \deeprane and use the Earth-mover distance as the loss function to minimize the gap between the predicted probability distribution over $100$ radar reflectivity classes. Then, we fine-tune the whole model using the loss function proposed in \cite{ko2022effective}, which is the negative average of approximated CSI scores defined as follows:
\begin{equation}
\mathcal{L}(C^{(t')}, \hat{C}^{(t')}) = -\frac{1}{2} \sum_{c \in \{\textsc{Heavy}, \textsc{Rain}\}} \frac{\widetilde{\text{TP}}_{c}}{\widetilde{\text{TP}}_{c} + \widetilde{\text{FP}}_{c} + \widetilde{\text{FN}}_{c}}, \label{eq:loss}
\end{equation}
where $\widetilde{\text{TP}}_{c}$, $\widetilde{\text{FP}}_{c}$, $\widetilde{\text{FN}}_{c}$ are differentiable approximations of the number of true positives, false positives, and false negatives for each precipitation class $c$. 
Refer to \cite{ko2022effective} for the details of the approximations and the pre-training step.
When training \method, we perform 
only the fine-tuning step using ground-based observations.

\section{Experiments}
\label{sec:exp}
\begin{table*}[t]
    \centering
    \caption{Predictive performances on the test set.
    \method performs best overall among the methods that use only the ground-based observations as inputs. \adv performs best overall among the methods that use ground-based observations and/or radar images as inputs.
    In each setting,
    the best score from the same input data is in \textbf{bold}. 
    }
    \scalebox{0.9}{
    \begin{tabular}{c|c|cc|cc|cc||cc|C{1cm}C{1cm}|C{0.75cm}C{0.75cm}}
        \toprule
        \multicolumn{2}{c|}{Input Data} & \multicolumn{6}{c|}{Ground-based Observations Only } 
        & \multicolumn{6}{c}{Ground-based Observations and Radar Images}
        \\
        \midrule
        \multirow{2}{*}{\makecell{Precipitation \\ level}} & \multirow{2}{*}{Lead time} 
        & \multicolumn{2}{c|}{\method} & \multicolumn{2}{c|}{LSTM} & \multicolumn{2}{c||}{Persistence} & \multicolumn{2}{c|}{\adv} & \multicolumn{2}{c|}{\deeprane + Kriging} & \multicolumn{2}{c}{\deeprane only} \\
        & & CSI & F1 & CSI & F1 & CSI & F1 & CSI & F1 & CSI & F1 & CSI & F1 \\
        \midrule
        \multirow{6}{*}{\makecell{\textsc{Heavy} \\ ($\geq$10mm/h)}} & $60$ minutes & 0.262 & 0.415 & \textbf{0.296} & \textbf{0.457} & 0.259 & 0.412 & \textbf{0.444} & \textbf{0.615} & 0.316 & 0.480 & 0.390 & 0.562 \\
        & $120$ minutes & 0.156 & 0.270 & \textbf{0.178} & \textbf{0.302} & 0.152 & 0.264 & \textbf{0.309} & \textbf{0.472} & 0.200 & 0.333 & 0.280 & 0.438 \\
        & $180$ minutes & 0.120 & 0.215 & \textbf{0.127} & \textbf{0.225} & 0.102 & 0.185 & \textbf{0.218} & \textbf{0.357} & 0.158 & 0.273 & 0.210 & 0.348 \\
        & $240$ minutes & \textbf{0.094} & \textbf{0.173} & 0.090 & 0.166 & 0.073 & 0.136 & 0.169 & 0.289 & 0.128 & 0.226 & \textbf{0.170} & \textbf{0.291} \\
        & $300$ minutes & \textbf{0.079} & \textbf{0.147} & 0.064 & 0.121 & 0.057 & 0.108 & \textbf{0.141} & \textbf{0.247} & 0.108 & 0.195 & 0.135 & 0.238 \\
        & $360$ minutes & \textbf{0.070} & \textbf{0.132} & 0.048 & 0.091 & 0.046 & 0.087 & 0.096 & 0.176 & 0.092 & 0.168 & \textbf{0.116} & \textbf{0.207} \\
        \midrule
        \multirow{6}{*}{\makecell{\textsc{Rain} \\ ($\geq$1mm/h)}} & $60$ minutes & \textbf{0.532} & \textbf{0.695} & 0.527 & 0.690 & 0.518 & 0.683 & \textbf{0.671} & \textbf{0.803} & 0.483 & 0.652 & 0.609 & 0.757 \\
        & $120$ minutes & \textbf{0.430} & \textbf{0.602} & 0.408 & 0.580 & 0.396 & 0.568 & \textbf{0.548} & \textbf{0.708} & 0.409 & 0.581 & 0.501 & 0.667 \\
        & $180$ minutes &\textbf{0.376} & \textbf{0.546} & 0.347 & 0.515 & 0.331 & 0.498 & \textbf{0.468} & \textbf{0.638} & 0.375 & 0.546 & 0.449 & 0.620 \\
        & $240$ minutes & \textbf{0.334} & \textbf{0.501} & 0.306 & 0.468 & 0.288 & 0.447 & \textbf{0.428} & \textbf{0.599} & 0.339 & 0.507 & 0.411 & 0.583 \\
        & $300$ minutes & \textbf{0.299} & \textbf{0.460} & 0.275 & 0.432 & 0.256 & 0.408 & \textbf{0.394} & \textbf{0.565} & 0.315 & 0.479 & 0.381 & 0.552 \\
        & $360$ minutes & \textbf{0.270} & \textbf{0.425} & 0.250 & 0.401 & 0.231 & 0.375 & \textbf{0.359} & \textbf{0.529} & 0.294 & 0.454 & 0.354 & 0.523 \\
        \midrule\midrule
        \multicolumn{2}{c|}{Average} & \textbf{0.252} & \textbf{0.382} & 0.243 & 0.371 & 0.226 & 0.348 & \textbf{0.354} & \textbf{0.500} & 0.268 & 0.408 & 0.334 & 0.482 \\
        \bottomrule
    \end{tabular}
    }
    \label{tab:q1_result}
\end{table*}

In this section, we review our experiments performed to answer the following questions:
\begin{enumerate}
    \item[Q1.] \textbf{Effectiveness of \adv and \method:} Is \adv more accurate than baseline methods? How does \method perform compared to baseline methods relying only on ground-based observations?
    \item[Q2.] \textbf{Ablation Study:} How much each ground-based meteorological feature contribute to the performance \adv? How does the self-attention block in \method and \adv affect their overall accuracy?
    \item[Q3.] \textbf{Further Analysis in Heavy Rainfall Cases:} How accurate is \adv for heavy rainfall cases with precipitation intensity levels exceeding $30$ mm/hr on the test set?
\end{enumerate}

\subsection{Experimental settings}

\subsubsection{Machine} We performed all experiments on a server with 512GB of RAM and eight RTX 8000 GPUs, each of which has 48GB of GPU memory.

\subsubsection{Datasets} For radar data, we used radar reflectivity images around South Korea that were measured every ten minutes from 2014 to 2020. Each radar image is 1468 x 1468 in size and has a 1 km x 1 km resolution.
For observations from ground weather stations, we used the data collected every 10 minutes from 2014 to 2020 from 714 automated weather stations (AWS) installed in South Korea.\footnote{While the data were collected every minute, we used 10\% of them.} %
The meteorological features collected from ground weather stations are (a) 9 features related to wind direction and speed, (b) a feature related to one-minute average temperature, (c) a binary feature related to precipitation, (d) 4 features related to cumulative precipitation in 15 minutes, 1 hour, 12 hours, and 24 hours, (e) a feature related to relative humidity, and (6) 2 features related to barometric pressure.

\subsubsection{Baseline approaches} %
As baseline approaches that use  ground-based observations and/or radar images, we used (a) \deeprane \cite{ko2022effective} and (b) \deeprane + Kriging. For the latter, we interpolated the ground-based observations to obtain grid-shaped data,\footnote{Since interpolation is expensive both in terms of computation and memory, we restricted the resolution to $4$ km $\times$ $4$ km and only used the last snapshot of ground-based observations (i.e., $O^{(t)}$).} using Kriging \cite{metheron1971theory}, and concatenated them to the input of \deeprane as additional channels. 
As baseline approaches that use only ground-based observations, we used (c) a vanilla LSTM model and (d) the persistence model. The LSTM model performs precipitation nowcasting based on the sequence of ground-based observations and the auxiliary information (spec., the location of the weather station, the observation date, the observation time, and the lead time), which are encoded as described in Section \ref{sec:method:integration}. Note that, compared to \method, the LSTM model uses all but radar images as its input. Also note that the LSTM model does not have the attention block, and thus once it is trained, it performs prediction independently for each region.
The persistence model is a simple heuristic that uses the current precipitation class in each region as its prediction in the region regardless of lead times.

\subsubsection{Experimental setup} 
For \deeprane and its training process, we used the open-source implementation provided by the authors.\footnote{\url{https://github.com/jihoonko/DeepRaNE}} For kriging, we used the PyKrige library.\red{\footnote{\url{https://geostat-framework.readthedocs.io/projects/pykrige/en/stable/}}} For the experimental setup, we mostly followed the protocol in \cite{ko2022effective}. For all experiments, we used the inputs from 2014 to 2018 for training, those in 2019 for validation, and those in 2020 for evaluation. In order to train models, we used the loss functions in Eq.~\eqref{eq:loss} and the Adam \cite{kingma2015adam} optimizer with a learning rate $2 \times 10^{-5}$. When pre-training \deeprane, we set the batch size to 20 and the number of training steps to $50,000$. For the other training processes, we set the batch size to 24 and the number of steps to $35,000$. For every $1,000$ steps, we evaluated the model\footnote{We measured the geometric mean of the two CSI scores for two precipitation classes, \textsc{Heavy} and \textsc{Rain}, at all lead times.} on the validation set and selected the trained model that performs best in the validation dataset. For \deeprane and the corresponding part of \adv, we set the number of initial hidden channels to $32$. For \method, we set the hidden dimension for LSTM cells to $64$ and the number of heads for the attention blocks to $4$.

\subsubsection{Evaluation metrics}

For evaluation, we used the CSI and F1 scores for two precipitation classes, \textsc{Heavy} ($\geq$ 10mm/hour) and \textsc{Rain} ($\geq$ 1mm/hour), at each lead time (i.e,. 1 to 6 hours). Their formula is given in Eqs.~\eqref{eq:csi} an \eqref{eq:f1}. They have been used widely for precipitation nowcasting \cite{agrawal2019machine,sonderby2020metnet,ko2022effective}.

\begin{table*}[t]
    \centering
    \caption{Ablation study results.
     The most crucial features among those in ground-based observations were cumulative precipitations, and removing the self-attention blocks from \adv was harmful to the overall performance. Recall that removing the self-attention blocks from \method was also harmful, as shown in Table~\ref{tab:q1_result}.
     In each setting, the best score is in \textbf{bold}.
     }
    \scalebox{0.9}{
    \setlength{\tabcolsep}{5pt}
    \begin{tabular}{c|c|cc|cc|cc|cc|cc|cc|cc|cc}
        \toprule
        \multirow{2}{*}{\makecell{Precipitation \\ level}} & \multirow{2}{*}{Lead time} & \multicolumn{2}{c|}{\adv} & \multicolumn{2}{c|}{\method-A} & \multicolumn{2}{c|}{\method-P} &   \multicolumn{2}{c|}{\method-W} & \multicolumn{2}{c|}{\method-T} &  \multicolumn{2}{c|}{\method-D} & \multicolumn{2}{c|}{\method-H} &  \multicolumn{2}{c}{\method-B}\\
        & & CSI & F1 & CSI & F1 & CSI & F1 & CSI & F1 & CSI & F1 & CSI & F1 & CSI & F1 & CSI & F1 \\
        \midrule
        \multirow{6}{*}{\makecell{\textsc{Heavy} \\ ($\geq$10mm/h)}} & $60$ minutes & 0.444 & 0.615 & \textbf{0.464} & \textbf{0.634} & 0.399 & 0.571 & 0.406 & 0.578 & 0.317 & 0.481 & 0.297 & 0.458 & 0.380 & 0.550 & 0.408 & 0.579 \\
        & $120$ minutes & \textbf{0.309} & \textbf{0.472} & 0.285 & 0.443 & 0.292 & 0.452 & 0.243 & 0.392 & 0.259 & 0.412 & 0.240 & 0.387 & 0.277 & 0.434 & 0.270 & 0.425 \\
        & $180$ minutes & \textbf{0.218} & 0.357 & \textbf{0.218} & \textbf{0.358} & 0.207 & 0.343 & 0.186 & 0.313 & 0.205 & 0.340 & 0.202 & 0.336 & 0.213 & 0.351 & 0.196 & 0.328 \\
        & $240$ minutes & \textbf{0.169} & \textbf{0.289} & 0.154 & 0.267 & 0.162 & 0.279 & 0.153 & 0.266 & 0.160 & 0.276 & 0.161 & 0.277 & 0.171 & 0.291 & 0.131 & 0.232 \\
        & $300$ minutes & \textbf{0.141} & \textbf{0.247} & 0.125 & 0.222 & 0.124 & 0.221 & 0.120 & 0.214 & 0.131 & 0.231 & 0.138 & 0.242 & 0.132 & 0.233 & 0.136 & 0.239 \\
        & $360$ minutes & 0.096 & 0.176 & 0.094 & 0.172 & 0.094 & 0.173 & 0.109 & 0.197 & 0.103 & 0.187 & \textbf{0.124} & \textbf{0.221} & 0.117 & 0.209 & 0.115 & 0.206 \\
        \midrule%
        \multirow{6}{*}{\makecell{\textsc{Rain} \\ ($\geq$1mm/h)}} & $60$ minutes & \textbf{0.671} & \textbf{0.803} & 0.670 & 0.802 & 0.657 & 0.793 & 0.623 & 0.768 & 0.588 & 0.740 & 0.645 & 0.784 & 0.609 & 0.757 & 0.630 & 0.773\\
        & $120$ minutes & 0.548 & 0.708 & \textbf{0.550} & \textbf{0.710} & 0.549 & 0.709 & 0.516 & 0.681 & 0.517 & 0.682 & 0.520 & 0.684 & 0.523 & 0.687 & 0.526 & 0.690\\
        & $180$ minutes & 0.468 & 0.638 & 0.477 & 0.646 & \textbf{0.488} & \textbf{0.656} & 0.451 & 0.622 & 0.466 & 0.635 & 0.460 & 0.630 & 0.468 & 0.637 & 0.463 & 0.633\\
        & $240$ minutes & 0.428 & 0.599 & 0.425 & 0.596 & \textbf{0.440} & \textbf{0.611} & 0.413 & 0.585 & 0.421 & 0.593 & 0.414 & 0.585 & 0.426 & 0.598 & 0.391 & 0.562\\
        & $300$ minutes & 0.394 & 0.565 & 0.390 & 0.561 & \textbf{0.404} & \textbf{0.576} & 0.366 & 0.535 & 0.390 & 0.561 & 0.380 & 0.550 & 0.390 & 0.561 & 0.393 & 0.564\\
        & $360$ minutes & 0.359 & 0.529 & 0.365 & 0.535 & \textbf{0.371} & \textbf{0.541} & 0.336 & 0.503 & 0.361 & 0.530 & 0.356 & 0.525 & 0.362 & 0.531 & 0.364 & 0.533\\
        \midrule\midrule
        \multicolumn{2}{c|}{Average} & \textbf{0.354} & \textbf{0.500} & 0.351 & 0.496 & 0.349 & 0.494 & 0.327 & 0.471 & 0.326 & 0.472 & 0.328 & 0.473 & 0.339 & 0.487 & 0.335 & 0.480\\
        \bottomrule
    \end{tabular}
    \label{tab:q2_result}
    }
\end{table*}

\subsection{Q1. Effectiveness of \adv and \method}
In order to demonstrate the effectiveness of \adv and \method, we compared the performances of them and the aforementioned baseline approaches.
The CSI and F1 scores are provided in Table \ref{tab:q1_result}. First, among the approaches that use only ground-based observations as the input, \method performed significantly better than the others, demonstrating the effectiveness of our proposed idea (i.e., using the attention blocks between LSTM cells). In addition, among the approaches that use ground-based observations and/or radar images, \adv performed best overall, achieving the $5.7\%$ better CSI scores on average than the others. 
Especially, for the precipitation class \textsc{Rain}, \adv performs best at all lead times. %
Notably, the performance of \deeprane + Kriging was even worse than \deeprane, while it outperformed all methods that use only ground-based observations as inputs. One potential reason is interpolation error induced by Kriging.

\subsection{Q2. Ablation Study}
We investigated the effect of the features in the ground-based observations and the self-attention block on the model's performance. To compare the performance, we compared \adv with the following variants, each of which was trained following the same training protocol of \adv:

\begin{enumerate}[leftmargin=*,label=(\alph*)]
\item \adv: the proposed method that uses all ground-based features.
\item \method-A: a variant without self-\textbf{a}ttention blocks in the architecture.
\item \method-W: a variant that uses only those related to \textbf{w}ind direction and speed among ground-based features.
\item \method-T: a variant that uses only those related to average \textbf{t}emperature among ground-based features.
\item \method-P: a variant that uses only those related to cumulative \textbf{p}recipitation among ground-based features.
\item \method-D: a variant that uses only  that related to precipitation \textbf{d}etection among ground-based features.
\item \method-H: a variant that uses only that related to relative \textbf{h}umidity among ground-based features.
\item \method-B: a variant that uses those related to \textbf{b}arometric pressure among ground-based features.
\end{enumerate}

We measured the CSI and F1 scores for each method, which are reported in Table \ref{tab:q2_result}. For the precipitation class \textsc{Heavy}, \adv achieved the best CSI and F1 scores on average, followed by \method-A and \method-P. 
Especially, \adv achieved the best CSI scores at all lead times except 1 and 3 hours.
While \adv was slightly outperformed by \method-P at most lead times for the precipitation class \textsc{Rain}, \adv still performed best overall when considering all $12$ settings, followed by \method-A and \method-P.
That is,  removing the self-attention blocks from \adv tended to be harmful, as removing them from \method was (compare \method and LSTM in Table~\ref{tab:q1_result}).
From the results, we also observed that, overall, the ground-based features for cumulative precipitation contributed most to the performance of \adv, followed by the those for relative humidity and barometric pressure.

\begin{figure*}
    \vspace{-2mm}
    \centering
    \includegraphics[width=\linewidth]{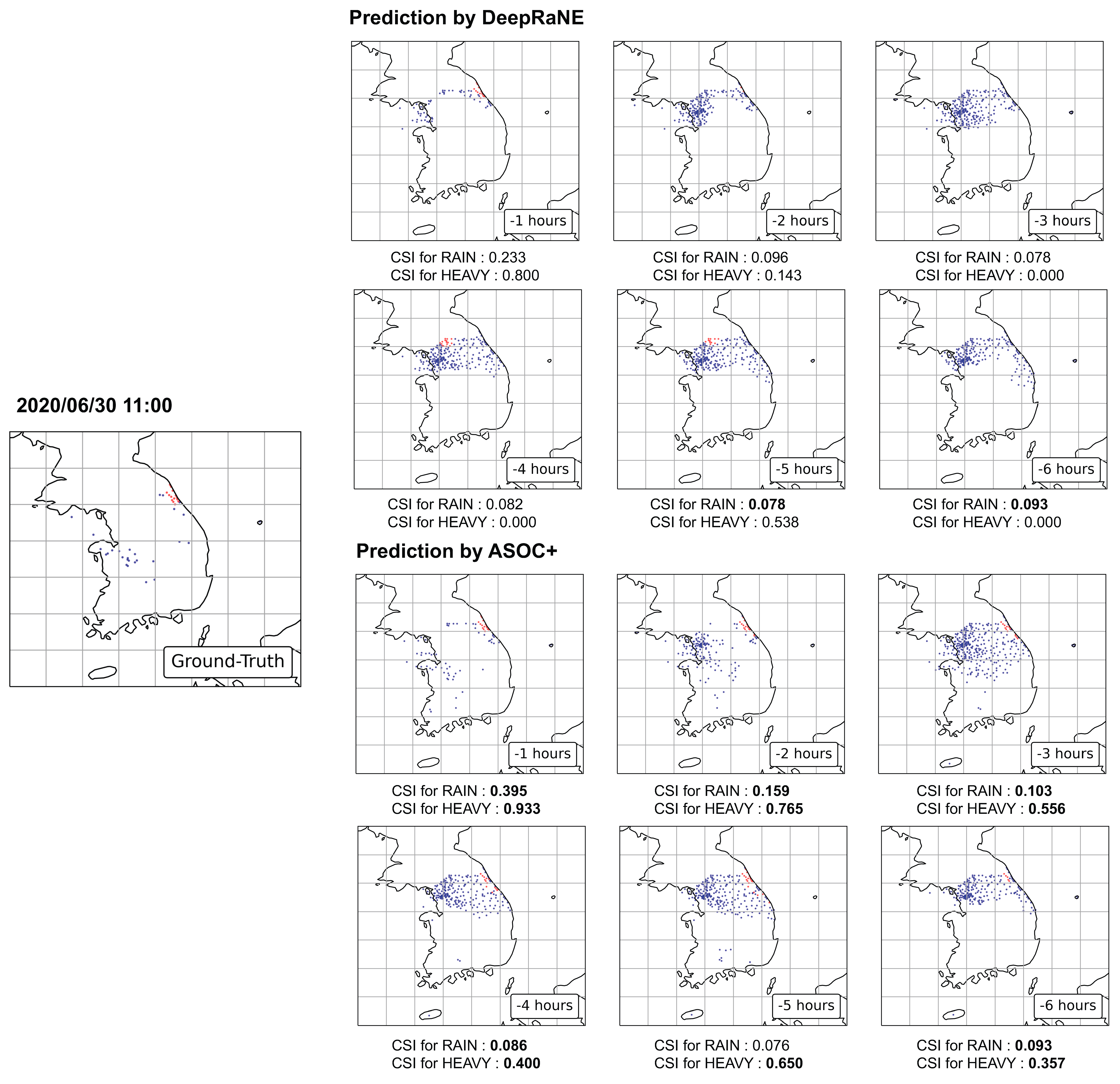}
    \caption{Comparison of predictions made by \adv and \deeprane for heavy rainfall cases (June 30, 2020 at 11 am). The  \textsc{Heavy} ($\geq 10$ mm/hr) cases and \textsc{Light} ($1$ - $10$ mm/hr) cases are marked in {\color{red} red} and {\color{blue} blue}, respectively. The largest figure on the left shows the ground-truth precipitation classes, and the others showed prediction results (spec., most probable classes) at different lead times.
    The CSI scores at each lead time are provided below the figures. Between the two CSI scores obtained by the two compared methods for each precipitation level at each lead time, the higher one is in \textbf{bold}. Note that the prediction results from the both methods become closer to the ground truth as the lead time reduces. Note that, although \deeprane fails to predict the locations of \textsc{Heavy} cases when the lead time is greater than $2$ hours, \adv predicts the locations quite accurately even when the lead time is $6$ hours.
}
    \label{fig:case}
\end{figure*}

\subsection{Q3. Further Analysis in Heavy Rainfall Cases}
We further analyzed the performance of \adv in $425$ cases in the test set where a precipitation intensity rate of 30 mm/hr or more was observed at one or more regions. On average, the CSI score for the precipitation classes \textsc{Heavy} ($\geq 10$ mm/hr) and \textsc{Rain} ($\geq 1$ mm/hr) were 0.234  and 0.465, respectively. Specifically, compared to \deeprane, \adv achieved 18.7\% (49.3\% at a 5-hr lead time) better CSI scores on average for \textsc{Heavy}  and 19.6\% (49.1\% at a 5-hr lead time) better F1 scores on average. Figure~\ref{fig:case} shows the detailed predictions provided by \adv and \deeprane at different lead times in one of the heavy-rainfall cases.

\section{Conclusion}
\label{sec:conclusion}
In this work, we proposed \method, a novel attentive and recurrent model for precipitation nowcasting using ground-based observations.
We designed \method to effectively utilize meteorological observations collected from multiple ground weather stations by allowing it to capture
temporal dynamics of the observations and the attentive relationships between them.
In addition, we proposed \adv, where
\method is combined with \deeprane, which is a state-of-the-art radar-image-based model.
Using ground-based observations (from $714$ ground weather stations) radar reflectivities collected around South Korea for seven years, we demonstrated the effectiveness of \method and \adv.
Specifically, we showed that \adv
improved the CSI score of predicting heavy (at least 10 mm/hr) and light (at least 1 mm/hr) rainfall events at 1-6 hr lead times by 5.7\%, compared to \deeprane. In addition, we measured the contribution of various features in ground-based observations to the performance of \adv and conducted a case study to further analyze the prediction results provided by \adv. 
For \textbf{reproducibility}, 
we made the source code used in the paper publicly available at \supplelink.

\section*{Acknowledgements}
{\small 
This work was funded by the Korea Meteorological Administration Research and Development Program ``Developing Intelligent Assistant Technology and Its Application for Weather Forecasting Process.'' (KMA2021-00123).
This work was supported by the Korea Meteorological Administration Research and Development Program
under Grant KMI2020-01010.
This work was supported by Institute of Information \& Communications Technology Planning \& Evaluation (IITP) grant funded by the Korea government(MSIT) (No.2019-0-00075, Artificial Intelligence Graduate School Program(KAIST)).}
\balance

\bibliographystyle{IEEEtran}
\bibliography{bib.bib}
\end{document}